\documentclass[prb,twocolumn,showpacs,amsmath,amssymb,aps]{revtex4}

\usepackage{graphicx}
\usepackage{graphics}
\usepackage{epsfig}
\usepackage{dcolumn}
\usepackage{bm}

\begin{document}

\title{Effects of signs in tunneling matrix elements on transmission zeros
and phase}
\author{Tae-Suk Kim$^{1}$ and S. Hershfield$^{2}$}
\affiliation{$^1$School of Physics, Seoul National University, 
  Seoul 151-747, Korea\\
 $^{2}$ Department of Physics, University of Florida, Gainesville FL 32611-8440 } 
\date{\today}

\begin{abstract}
The effect 
of the signs in the tunneling matrix elements on the transmission zeros and 
the transmission phase in transport through a quantum dot is studied.
The existence of transmission zeros is determined by both the relative signs
and the strength of the tunneling matrix elements
for two neighboring energy levels of a dot.
The experimentally observed oscillating behavior of the transmission phase
over several Coulomb peaks can be explained by a uniform distribution of
the relative signs. Based on a simple model of a quantum dot, 
we present a possible scenario which can give such uniform signs over several 
conductance peaks. We suggest that the location of the transmission zeros
can be identified by inspecting the Fano interference pattern 
in the linear response
conductance of a closed Aharonov-Bohm (AB) interferometer with an embedded quantum dot
as a function of the number of electrons in the dot and the AB flux.

\end{abstract}
\maketitle

\section{Introduction}
 Landauer's scattering theory has been very successful in understanding 
transport phenomena in the mesoscopic systems.
Electrons, when scattered, experience a phase shift in their
wave functions and this phase shift (Friedel phase) is stored in
the scattering matrix. 
Though the transmission amplitude 
contains information about both its magnitude and phase (transmission phase),  
the conductance measures only the absolute value of the transmission amplitude.
Recently a number of experiments have 
measured the phase shift experienced by  electrons
passing through a quantum dot, using Aharonov-Bohm (AB) interferometers
\cite{closering,openring1,openring2,openring3}.
The measured phase is not the Friedel phase as we know in bulk systems, but 
the transmission phase which is the argument of the transmission amplitude
through a quantum dot. 
In a closed AB interferometer, the measured phase is constrained to be
either $0$ or $\pi$ as a 
consequence of the Onsager relations. 
The phase is observed to jump by $\pi$ at the conductance peaks 
and remains in phase between successive Coulomb peaks\cite{closering}.

 In order to observe the phase evolution in a quantum dot, 
an open AB interferometer similar to the double-slit experiments
was devised by Schuster, et. al\cite{openring1}. 
The measured phase evolves smoothly by $\pi$ over the Coulomb peaks, 
which can be expected for resonant tunneling.
The transmission phase jumps by $\pi$ somewhere in between two neighboring peaks
and remains in phase between successive Coulomb peaks. 
This unexpected sudden phase jump can be understood in terms of the transmission zeros.
\cite{lee,buttiker,tskim}
Oscillating behavior of the phase -- a smooth increase over the Coulomb peaks 
and a sudden drop in between -- is observed over several successive Coulomb peaks.

 In this paper we study theoretically 
the effects of the signs of the tunneling matrix elements
between a quantum dot and two leads
on the transmission probability $T_Q$ of a quantum dot.
Silva, Oreg, and Gefen have recently shown that the
signs of the tunneling matrix elements play a crucial role
in determining the behavior of the transmission probability and phase
by studying a quantum dot with two resonant levels.
\cite{signphase}
A number of other works have either explicitly or implicitly 
addressed the role of the relative sign of the tunneling matrix
elements.\cite{gefenreview}
In this paper, we consider quantum dots with many levels
and consider several possible distributions of the signs 
in the tunneling matrix elements for each energy 
level in a quantum dot.
The effects 
of a given distribution of the signs on the transmission phase and zeros
is studied.
Although the distribution of the signs in the tunneling 
matrix elements is computed for one simple model, 
the emphasis here is on understanding the general relationship between the
transmission phase and the relative signs of the matrix elements.
This can be used, for example, to work backward from experiment to
determine the relative signs of the tunneling matrix elements.
Not all distributions of the relative signs of the tunneling matrix
elements may be physically realizable.
It is still an open question why the phase jump 
or the transmission zeros show up over the energy range of several Coulomb peaks.
Transmission zeros can arise from the interference between the possible 
current paths through the multiple energy levels in a dot,
and can explain the abrupt phase jump by $\pi$.

 We find from our model study that 
the oscillating behavior in the transmission phase is a result of 
having
the same relative signs in the tunneling matrix elements. 
Successive transmission zeros in some energy interval
arise when the relative signs
in the left and right tunneling matrix elements are the same 
for all the energy levels lying inside that interval.  
The well-studied $t$ stub\cite{buttiker,tskim,lent} belongs to this category.
The transmission zeros are absent from the interested energy interval
when the relative signs oscillate from level to level.
In this case, the transmission phase steadily increases by $\pi$ 
over every peak in $T_Q$. 
The double-barrier well\cite{buttiker,tskim,lent}
 shows this type of structure in the transmission amplitude. 
These results are in agreement with those found earlier in 
two level quantum dots.\cite{signphase}

Based on studying several different cases for the relative signs
of the tunneling matrix elements, we are able to deduce simple rules
governing the presence or absence of transmission zeros
when the tunneling matrix elements are much smaller than the interlevel energy spacing.
In addition
depending on the distribution of signs, the transmission probability
in an AB interferometer with an embedded dot takes on unique shape,
especially as a function of the Aharonov-Bohm flux. 
Thus, 
complementary to the direct measurements of the transmission phase 
in the open AB interferometer, we suggest that 
the position of transmission zeros of the quantum dot
can be located 
by inspecting the Fano interference pattern in the linear response conductance.

 The rest of this paper is organized as follows. In Sec.~\ref{sect_model}, 
the model Hamiltonians and the formalism are described. 
We present our results of our model studies in Sec.~\ref{sect_results}
and discuss a particular model for the relative signs
of the matrix elements in Sec.~\ref{sect_simple}. 
Our study is summarized and concluded
in Sec.~\ref{sect_sum}. 
In the Appendix, the case of a complex Fano asymmetry factor in an AB 
ring geometry is discussed.

\section{Model and formalism \label{sect_model}}
 To study the effect of the signs of the tunneling matrix elements between 
a quantum dot and two leads on the transmission probability and phase, 
we consider a quantum dot which is modeled by discrete single-particle
energy levels and is connected to two metallic leads by tunneling.
The metallic leads can accommodate only one transport channel.
The model Hamiltonian is given by the equation, $H = H_0 + H_1$, where
\begin{eqnarray}
H_0 &=& \sum_{p=L,R} \sum_{k\alpha} \epsilon_{pk} c_{pk\alpha}^{\dag} c_{pk\alpha}
       + \sum_{i\alpha} E_i d_{i\alpha}^{\dag} d_{i\alpha}, \\
H_1 &=& \sum_{pk} \sum_{i\alpha}  V_{pi} c_{pk\alpha}^{\dag} d_{i\alpha} + H.c.
\end{eqnarray}
$H_0$ is the Hamiltonian for the isolated system of two leads and a quantum dot, and 
$H_1$ is the tunneling Hamiltonian between a dot and two leads.
$E_i = \epsilon_i - E_g$ is the energy level of a dot which is shifted
by the gate voltage ($E_g$ is proportional to the gate voltage). 
The two metallic leads are conveniently named the left and right leads.  
The transmission amplitude for an electron moving from the left to right lead
through the mesoscopic system can be expressed in terms of the mixed Green's function,
\begin{eqnarray}
t_{RL}(\epsilon)
 &=& -2 \frac{G_{RL}^r(\epsilon)}{\pi\sqrt{N_L N_R}}, 
\end{eqnarray}
where $N_{L,R}$ is the density of states in the left (right) lead
at the Fermi level. 
The mixed Green's function $G_{RL}$ is defined by 
\begin{eqnarray}
i\hbar G_{RL} (t,t') 
 &=& \sum_{kk'} \langle T c_{Rk}(t) c_{Lk'}^{\dag}(t') \rangle.
\end{eqnarray}
The transmission amplitude through a quantum dot can be expressed 
in terms of the Green's function of the dot.
\begin{eqnarray}
\label{tampqdot}
\frac{t_{Q} (\epsilon)}{ 2\pi \sqrt{ N_L N_R} } 
 &=& \sum_{ij} V_{Ri} D_{ij}^{r} (\epsilon) V_{jL}.
\end{eqnarray}
This transmission amplitude leads to the familiar expression of 
the transmission probability $T(\epsilon) = |t_{RL}|^2$.
\begin{eqnarray}
\label{tprobqdot}
T(\epsilon) 
 &=& 4 \mbox{Tr} \{ {\bf \Gamma}_L {\bf D}^{r}(\epsilon) {\bf \Gamma}_R 
                    {\bf D}^{a} (\epsilon) \}.
\end{eqnarray}
${\bf D}^{r,a}$ is the retarded (advanced) Green's function of the dot.
The Green's function of 
a noninteracting dot is given by the expression, 
\begin{eqnarray}
{\bf D}^{r,a} (\epsilon) 
 &=& [ \epsilon {\bf I} - {\bf E} \pm i {\bf \Gamma} ]^{-1},
\end{eqnarray}
where
$E_{ij} = E_i \delta_{ij}$ and 
$\Gamma_{ij} = \pi N_L V_{iL} V_{Lj} + \pi N_R V_{iR} V_{Rj}$.
For this noninteracting dot, the transmission zeros
of $t_Q$ are sensitive to the relative phases of $V_{Li}$ and $V_{Ri}$, 
since $t_Q$ is the algebraic sum of $D_{ij}$ weighted by the tunneling 
matrix elements.

 An Aharonov-Bohm interferometer is made by introducing 
direct tunneling between the two leads,
\begin{eqnarray}  
H_2 &=& \sum_{k,k'\alpha} T_{LR} c_{Lk\alpha}^{\dag} c_{Rk'\alpha} + H.c.
\end{eqnarray} 
The transmission amplitude through the AB interferometer with an embedded 
quantum dot is given by the equation
\begin{eqnarray}
\label{tampring}
\frac{t_{AB} (\epsilon)}{ 2\pi \sqrt{ N_L N_R} } 
 &=& \frac{T_{RL}}{1+\gamma} 
  + \sum_{ij} \tilde{V}_{Ri} D_{ij}^{r} (\epsilon) \tilde{V}_{jL}.
\end{eqnarray}
Here $\gamma = \pi^2 N_L N_R |T_{LR}|^2$ is the dimensionless measure of 
the direct tunneling rate between the two leads and is related to the direct tunneling 
probability by $T_0=4\gamma/(1+\gamma)^2$. The tunneling matrix elements
 with a tilde
can be considered as being renormalized by the direct tunneling:
$\tilde{V}_{iL} = [V_{iL} - i\pi V_{iR} N_R T_{RL}]/(1+\gamma)$ and
$\tilde{V}_{Ri} = [V_{Ri} - i\pi T_{RL} N_L V_{Li}]/(1+\gamma)$. 
The Green's function of the dot is also modified by the direct tunneling
and the dot's self-energy is given by the expression:
\begin{eqnarray}
\Sigma_{d,ij}^{r} 
 &=& - i \sum_{p=L,R} \frac{\pi N_p V_{ip}V_{pj}}{1+\gamma} \nonumber\\
 && - \frac{\pi^2 N_L N_R} {1+\gamma} \sum_{p=L,R} V_{ip} T_{p\bar{p}} V_{\bar{p}j}.
\end{eqnarray}
Here the notation $\bar{L} = R$ and $\bar{R} = L$ is used. 
The effect of the Aharonov-Bohm flux $\Phi$ threading through the AB interferometer
can be taken into consideration by introducing the AB phase 
$\phi = 2\pi \Phi/\Phi_0$ into the tunneling matrix elements
as $V_{Ri} V_{iL}T_{LR} = |V_{Ri} V_{iL}T_{LR}| e^{i\phi}$, where
$\Phi_0 = hc/e$ is the flux quantum.

 The behavior of the transmission phase depends sensitively 
on the relative phases 
of the tunneling matrix elements. In the time-reversal symmetric case, an electron's
wave functions can be chosen to be real. 
For a real barrier potential, the tunneling matrix elements $V_{pi}$
($p=L,R$) are thus all real. 
The signs of tunneling matrix elements can be introduced
by the relations:
\begin{eqnarray}
V_{Li} &=& s_{Li} |V_{Li}|, ~~~ V_{Ri} ~=~ s_{Ri} |V_{Ri}|.
\end{eqnarray}
Transforming
the electron operator in the dot via
\begin{eqnarray}
d_{i\alpha} &\to& s_{Li} d_{i\alpha},
\end{eqnarray}
all the phases are ascribed to the right tunneling matrices. 
The left tunneling matrices are all positive, $V_{Li} > 0$ for all $i$'s,
and  
\begin{eqnarray}
V_{Ri} ~=~ s_i ~ |V_{Ri}|, ~~~ s_i = s_{Li} s_{Ri}, 
\end{eqnarray}
where $s_i = \pm 1$. We have no information about the distribution of $s_i$
at this point. 
Though the tunneling matrix elements are all assumed to be real, 
the conclusion remains the same as long as 
the relative phases between the left and right tunneling matrix elements
are real. An overall random phase in the left and right tunneling
matrices does not change the physics.

\section{Transmission phase and distribution of relative signs in tunneling 
matrix elements \label{sect_results}}
 In this section, we consider several possible distributions of 
the relative signs in the left and right tunneling matrix elements
 and study the behavior of the transmission zeros and phase.
We find the general rule which governs the relationship between 
the signs and the transmission phase. 
The transmission probability through a dot 
takes on many different shapes depending
on the distribution of the signs.
A closed or unitary AB interferometer with an embedded quantum dot 
can thus be an auxiliary experimental probe to measure the transmission zeros 
and phase.

 Before presenting our results, we briefly comment on the validity 
of our single-particle approach to a quantum dot. 
 The transmission probability through a quantum dot in a two-terminal 
configuration is the sum of the Green's functions of the dot
weighted by the tunneling matrix elements [see Eq.~(\ref{tampqdot})]. 
The poles in the Green's functions are determined by the zeros of the 
determinant of $[D^r]^{-1}$. 
Though the position of the zeros in the determinant 
depends on the relative signs of the tunneling matrix elements,  
the existence of zeros in the determinant is not sensitive to 
the relative signs.
On the other hand, the existence of the transmission zeros in $t_Q$ 
is sensitively determined by the relative signs of the tunneling matrix elements. 
At the level of the Hartree
approximation, the main effect of the Coulomb interaction is to
to make discrete poles
which are separated by the Coulomb interaction plus the difference 
in the single-particle energy spectrum. The level broadening is 
mainly determined by the tunneling into two leads.  
Since Eq.~(\ref{tampqdot}) is valid at the level of the Hartree approximation, 
we believe the transmission zeros 
are still determined by the relative signs of the tunneling matrix elements,
although we do not consider this case explicitly here.

 We would like to emphasize that our consideration of the distributions
of the relative signs in the left and right tunneling matrix elements
do have physical realizations.
To obtain the information about the tunneling matrix 
elements requires the solution of the Schr\"{o}dinger equation for the 
entire structure of a quantum dot and two leads.
Simplifying the system to the one-dimensional problem, there have been
some approaches to computing the transmission amplitude of 
the quantum dot system. Modeling the quantum dot by a double-barrier well (DBW)
or t stub, plane waves are matched at the boundaries between the quantum dot 
and the two leads.
The structure of the transmission amplitude for the DBW (t stub)
is completely equivalent to the case of the oscillating (uniform) 
distribution of the relative signs, as will be discussed below. 
The stepwise distribution of the relative signs is also possible
for the double-barrier t stub system.

 For our model study, we assume the quantum dot is described by 
a uniform single-particle energy spectrum with energy spacing $\Delta$.
The level spacing implicitly
includes the effect of the Coulomb interaction 
at the Hartree approximation level. 
We keep a finite number of energy levels with 
$\epsilon_i = i\Delta$ where the integer $i$ is constrained to 
$-N \leq i \leq N$ and $N = 50$ is chosen in our plots.
The transmission probabilities $T_Q(0)$ of a quantum dot,
$T_{AB}(0)$ of an AB interferometer at the Fermi energy $\epsilon=0$,
and the transmission phase $\theta$ of a quantum dot are computed 
as a function of the energy shift $E_g$ by the gate voltage. 
In most of the presented figures, $E_g$ is varied from $-4\Delta$ 
to $4\Delta$. This energy range covers 9 discrete energy levels of the dot. 
Unless otherwise noted,
we use the same set of the tunneling matrix elements
to emphasize the effects of $s_i$'s.
The values of $\Gamma_{Li}/\Delta$ ($\Gamma_{Ri}/\Delta$)
for $i = -N,-N+1, \cdots, N$ 
are chosen randomly between $0.01$ ($0.03$) and $0.02$ ($0.06$), respectively,
where $\Gamma_{pi} = \pi N_p V_{iL}V_{Li}$ for $p=L,R$. 
When the quantum dot is embedded into one arm of an AB interferometer, 
the effect of the AB flux is taken into account by introducing 
an AB phase $\phi$ into the tunneling matrices.   
Though the energy spectrum and the tunneling matrix elements of a dot
are modified by magnetic fields (see Sec.~\ref{sect_simple}),
our presented figures are computed using a rigid energy spectrum and 
rigid tunneling matrix elements.

\begin{figure}
\includegraphics{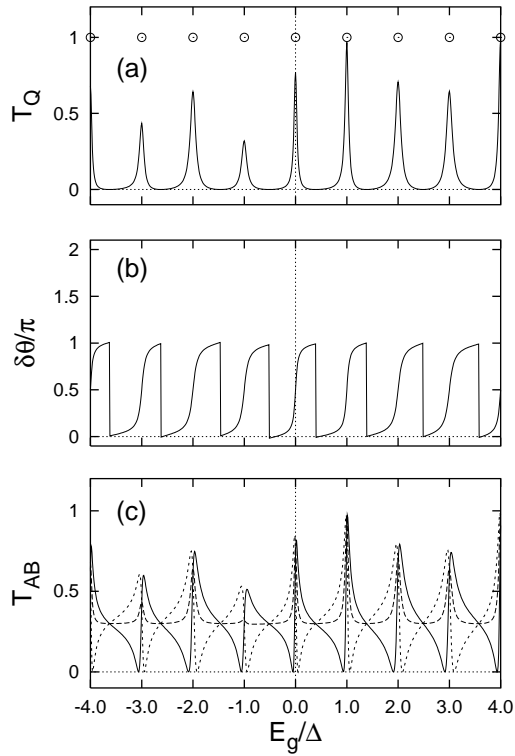} 
\caption{Uniform distribution of relative signs ($s_i = 1$). 
(a) The transmission probability $T_Q$ of a quantum dot.  
Open circles denote the relative signs of the tunneling matrix elements.
$s_i = \pm 1$ when an open circle is located at $1 (0)$ on the $y$-direction, 
respectively.
(b) The variation of the transmission phase $\delta\theta$. 
(c) When the quantum dot is embedded into the Aharonov-Bohm interferometer,
the Fano interference pattern shows up in the transmission probability 
$T_{AB}$. The AB phase ($\phi$) dependence of $T_{AB}$ is displayed ($T_0 = 0.3$): 
solid line ($\phi = 0^{\circ}$), long dashed line ($\phi = 90^{\circ}$), 
and short dashed line ($\phi = 180^{\circ}$).   
\label{tstub}}
\end{figure}

 To begin we consider the case when all the relative signs are equal,
i.e., $s_i = 1$ for all $i$'s. The transmission probability $T_Q$ 
 of the quantum dot is displayed for this case in Fig.~\ref{tstub} (a). 
$T_Q$ has a repeating sequence of zero-pole pairs.
The transmission probability is peaked
whenever the Fermi energy is aligned with one of discrete energy levels 
in the dot. 
$T_Q$ is zero at the transmission zeros, which 
lie in between two neighboring transmission peaks or poles. 
The change in the transmission phase, $\delta \theta$,
is displayed in Fig.~\ref{tstub} (b). $\delta\theta$ increases by $\pi$ 
over every transmission peak and jumps by $\pi$ at transmission zeros.
$\delta\theta$ shows an oscillating behavior: an increase over poles and
a drop at zeros.  
 The transport properties of this quantum dot system are very similar to 
those of well-studied $t$ stub system. \cite{buttiker,tskim,lent}

 When the quantum dot is embedded in an AB interferometer, the transmission
probability $T_{AB}$ shows Fano interference\cite{Fano}
between the direct tunneling and the resonant path through the dot.
As shown in Fig.~\ref{tstub} (c), a repeating sequence of zero-pole pairs 
(solid line) 
still remains and additional zeros are not generated by the destructive 
Fano interference.
The asymmetric Fano resonance shape in $T_{AB}$ for each peak 
are equivalent or ``in phase" in the sense that 
they are modulated in the same way by the AB flux. 
$T_{AB}(\phi=0^{\circ})$ has an $N$-shape structure,
and the shape of $T_{AB}$ is transformed into an inverse $N$-shape
when $\phi = 180^{\circ}$.  
The transmission zeros of $T_{AB}$ lie on the real-energy axis when 
$\phi = n\pi$, where $n$ is an integer.
 For other values of the AB phase $\phi$, 
the transmission zeros become complex.
Due to the Onsager relation, $T_{AB} (-\phi)$ is equal to $T_{AB} (\phi)$.

In recent measurements of the transmission phase, $\theta$,
of a quantum dot in an open AB interferometer, oscillating behavior 
of $\theta$ was observed over several Coulomb peaks.  
Within our model, these experimental features can be 
explained when the relative signs
of the tunneling matrix elements are all equal.

\begin{figure}
\includegraphics{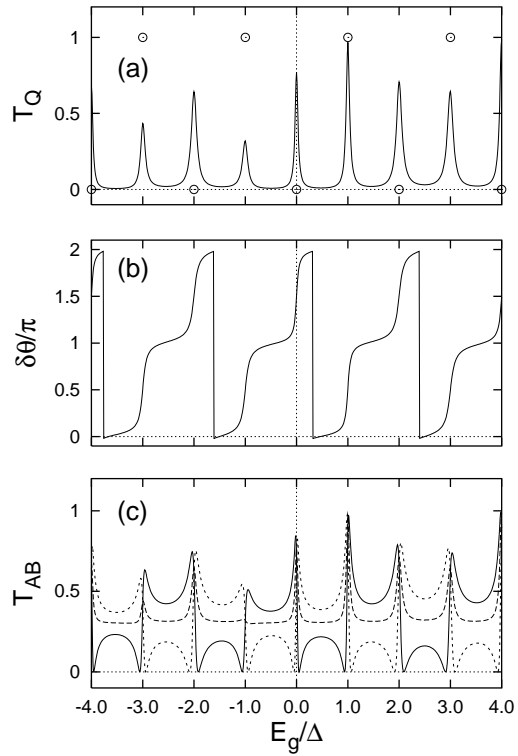} 
\caption{Oscillating distribution of relative signs
($s_i = (-1)^{i+1}$). The integer $i$ labels the single-particle energy levels 
in a dot. 
(a) $T_Q$ at the Fermi energy. The meaning of open circles are explained
in the caption of Fig.~\ref{tstub}.
(b) The transmission phase. For display, $\theta$ is folded into 
the range $[0,2\pi]$ so that the apparent jump of $\theta$ by $2\pi$
should be understood as continuous line. 
(c) $T_{AB}$ in an AB ring. Lines are the same as in Fig.~\ref{tstub}. 
  \label{dbw}}
\end{figure}

 When the relative signs of the tunneling matrix elements are 
oscillating from level to level,
$s_i = (-1)^{i+1}$,
the structure of transmission zeros is drastically modified.
There are no transmission zeros in $T_Q$, but instead $T_Q$ has
a series of transmission poles. 
Since no transmission zeros exist in $T_Q$, the transmission phase
steadily increases by $\pi$ as the energy levels of the dot are scanned
by varying $E_g$ 
through the Fermi level of the leads. 
The behavior of this quantum dot system is very similar to a double-barrier
well and double-barrier resonant tunneling systems.\cite{buttiker,tskim,lent}

 When this quantum dot is inserted into an AB interferometer,
destructive Fano interference generates transmission zeros 
in $T_{AB}$ in an energy interval of $E_g$ where $T_Q$ has no zeros.
 $T_{AB}$ has a repeating sequence of 
zero-zero-pole-pole structure. 
Although the difference in the $T_Q$'s is not significant
between the two distributions of 
$s_i$'s in Fig.~\ref{tstub}(a) and Fig.~\ref{dbw}(a),
the shapes of $T_{AB}$ are quite different. 
In this case
$T_{AB}$ consists of a repeating sequence of the structure of $M$-shape and $W$-shape
when $\phi=0^{\circ}$. The $M (W)$-shape is transformed into the $W (M)$-shape 
when $\phi = 180^{\circ}$. 
The difference in 
Fig.~\ref{tstub}(c) and Fig.~\ref{dbw}(c)
suggests that these two types of distributions in $s_i$'s 
can be distinguished by measuring the linear response conductance $G_{AB}$
in the AB interferometer with an embedded quantum dot 
and by inspecting the shapes of $G_{AB}$ as a function of the gate voltage
and magnetic field.

\begin{figure}
\includegraphics{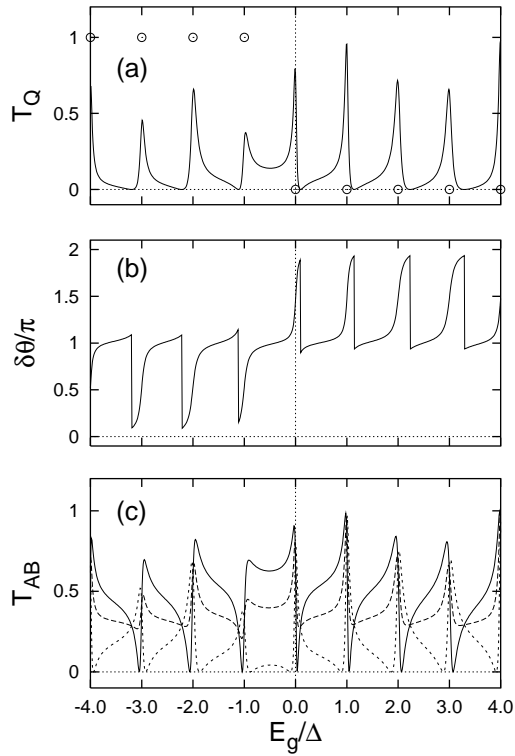} 
\caption{Stepwise distribution of relative signs 
($s_i = 1$ for $i \leq -1$ and $s_i = -1$ for $i \geq 0$).
(a) The asymmetrical lineshape in $T_Q$ arises from the combined effects 
of the stepwise distribution of relative signs
and the interference between multiple energy levels.
(b) $\delta\theta$. One transmission zero is removed in the boundary energy 
range $-1 < E_g/\Delta < 0$.
(c) $T_{AB}$ in an AB ring.  Lines are the same as in Fig. 1.
\label{step}}
\end{figure}

 We further explore other possible distribution of $s_i$'s.  
The behavior of $T_{Q,AB}$ and $\theta$ are displayed in Fig.~\ref{step}
when $s_i = 1$ for $i\leq -1$ and $s_i = -1$ for $i \geq 0$.
In this stepwise distribution of the relative signs,
the transmission phase $\delta\theta$ show the expected oscillating behavior
due to transmission zeros
[see Fig.~\ref{tstub}(b)]
in the two energy ranges of uniform relative signs 
[see Fig.~\ref{step}(b)].
However, one transmission zero is removed in the energy interval which lies in between the
two energy levels whose signs ($s_i$) are opposite.
Removal of one zero leads to a continuous increase of $\delta \theta$ 
by $\pi$ over each energy level ($E_g = - \Delta$ and $E_g = 0$ in our 
model calculation).
As displayed in Fig.~\ref{step}(a), 
the shape of $T_Q$ is now asymmetrical with respect to 
each peak position.
Furthermore, the shape of $T_Q$ in both energy regions of uniform signs
is different by a ``phase" of $\pi$ 
in the sense that the AB phase $\pi$ can change the shape of $T_{AB}$
in one region into that of $T_{AB}$ in the other, 
and vice versa [See Fig.~\ref{step}(c)].
Obviously this phase change $\pi$ in $T_Q$ stems from the sign flip in the tunneling
matrix elements. 
The absence of one transmission zero in $T_Q$ 
is much more apparent in $T_{AB}$.

\begin{figure}
\includegraphics{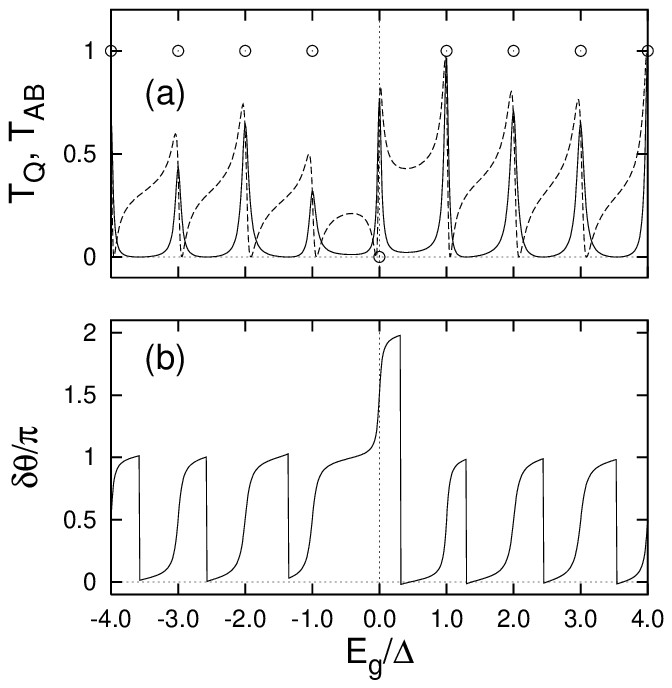}
\includegraphics{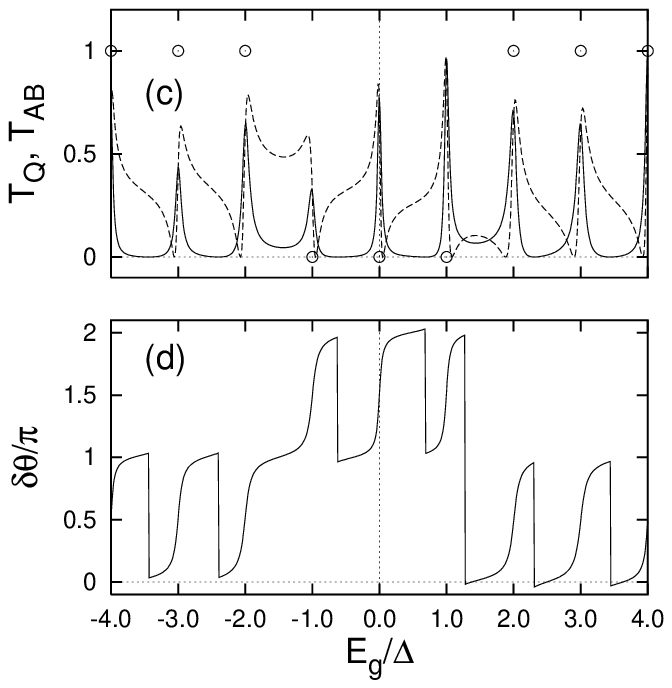}
\caption{Uniform relative signs ($s_i = 1$) 
except for one flipped sign ($s_i = -1$).
(a) and (c): Transmission probabilities $T_{Q}$ (solid line) 
and $T_{AB}$ (dashed line). 
(b) and (d): The transmission phase of a dot.
The jump by $2\pi$ should be understood as a continuous line
because the phase is defined to be between $0$ and $2\pi$.
\label{interval}}
\end{figure}

 We now consider the cases that all the energy levels have the same relative signs
except for one energy interval with the flipped relative signs. 
The energy levels belonging to this flipped energy interval
have the opposite relative signs, $s=-1$, compared to all the other
levels with $s=1$. 
In Fig.~\ref{interval} (a) and (b), we study $T_{Q,AB}$ and $\delta\theta$
when the flipped energy interval consists of only one level, $i=0$.
Transmission zeros are removed only in the energy range lying in between 
the two levels with the opposite signs. Since no zeros exist for 
$-\Delta \leq E_g \leq \Delta$, the transmission phase increases by $\pi$ 
over each level $i=-1,0,1$. 
The shape of $T_Q$ is similar to the case of uniform
distribution of $s_i$'s 
except for the removal of transmission zeros.
On the other hand $T_{AB}$ ($\phi=0$) is significantly 
modified in the flipped energy interval. 
As the number of the flipped levels is increased,
transmission zeros are removed only in the energy range lying in between 
the
two levels with opposite signs or in the interface energy range 
between the flipped interval and the rest. 
Transmission zeros and the corresponding phase jumps 
persist in between two levels with the same relative signs.
The case when the
the flipped interval consists of three levels is studied
in Fig.~\ref{interval} (c) and (d). 
Examination of $T_{AB}$ reveals that the ``phase" of $T_{AB}$
is different between the inside and outside of the flipped interval. 
The interval acts to shift the phase by $\pi$ at the two boundaries,
and the shape of $T_{AB}$ is identical on both sides of the flipped interval.

 When the line-broadening is much smaller 
than the interlevel spacing,
we can deduce from these results the following rules 
for the behavior of the transmission zeros. 
 When the relative signs are opposite for the two neighboring energy levels, 
the transmission zero is absent in the energy interval bounded by two levels.   
 When the relative signs are the same for two neighboring energy levels,
one transmission zero is generated inside the energy interval bounded by two levels.
This conclusion can be modified when one of the linewidth matrix elements
is by far larger than the others as will be shown below.

\begin{figure}
\includegraphics{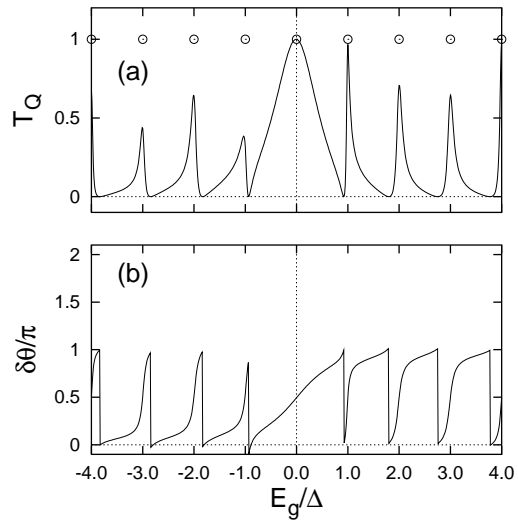} 
\caption{(a) Transmission probability $T_Q$ and (b) transmission phase for a dot
when the relative signs are uniform and one energy level is much more
strongly coupled to two leads than the other energy levels. 
Even with the same relative signs for all energy levels, the lineshape is
asymmetrical.  Model parameters are the same as in the previous figures 
except that one of the line broadening parameters, 
$\Gamma_{L,R}$ for $i=0$, is chosen to be large: $\Gamma_{L0}=\Gamma_{R0}
= \Delta/2$. \label{largestub}}
\end{figure}

 The lineshape in $T_Q(E_g)$ can be asymmetrical depending on the relative
signs of the tunneling matrix elements and their magnitude. 
One example of an asymmetrical lineshape is already displayed
in Fig.~\ref{step} for the stepwise distribution of relative signs. 
Another way to make the lineshape of $T_Q$ asymmetrical is to make 
one energy level more strongly coupled to the two leads than
the other levels.  
This strongly coupled level is broadened and 
 can act as an effective continuum band to other narrow levels. 
The interference between the broad level and
the narrow levels leads to the asymmetrical Fano lineshape. 
A typical example is displayed in Fig.~\ref{largestub}
when the relative signs in the tunneling matrix elements are equal
for all energy levels.
Although the lineshape in $T_Q$ is modified compared to 
Fig.~\ref{tstub}, the transmission zeros for the uniform distribution of 
signs persist.

\begin{figure}
\includegraphics{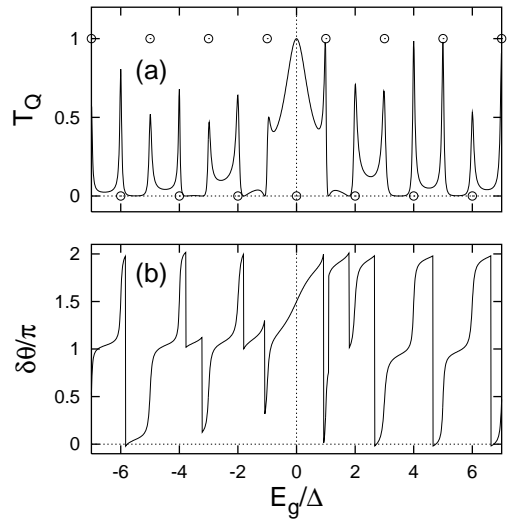} 
\caption{(a) Transmission probability $T_Q$ and (b) transmission phase for a dot
when the relative signs are oscillating from level to level 
 and one energy level is much more
strongly coupled to two leads than the other energy levels. 
Model parameters are the same as in Fig.~\ref{largestub}.   
Note the far-reaching effect of the strong coupling for one level
on the other levels. 
The jump by $2\pi$ in $\delta\theta$ should be understood as a continuous line.
\label{largedbw}}
\end{figure}

 In addition to the asymmetrical lineshape, 
a more dramatic effect of Fano interference (see Fig.~\ref{largedbw}) can occur in
the transmission zeros 
when the relative signs are oscillating from one level to the next. 
As shown in Fig.~\ref{dbw},
the transmission zeros are absent for the oscillating distribution of signs
when the linewidth is much smaller than the interlevel spacing.
Surprisingly, new transmission zeros are generated due to Fano interference
between the broad energy level and the other narrow levels.
Note that the strong coupling of one level to the leads is very far-reaching
and has the same effect on the structure of the transmission amplitude
as the AB interferometer.

\section{Model calculation of relative signs\label{sect_simple}}
 In the previous section, we studied the transmission zeros and phase
of a quantum dot for several possible distributions of the relative signs 
in the tunneling matrix elements.
A uniform distribution of signs is consistent with 
recent experimental results. 
In this section we present a simple scenario which can lead to a
uniform distribution of signs.

 The single-particle energy-level spectrum in a quantum dot has been inferred 
by monitoring the position of the Coulomb-regulated conductance peaks
as a function of magnetic field\cite{mceuen}. 
After subtracting the Coulomb interaction from the difference between two 
conductance peaks, the single-particle energy spectrum was inferred and 
well-described by the Fock-Darwin energy levels 
for some range of magnetic fields. 
In this section we show that 
the Fock-Darwin energy spectrum\cite{fock, darwin} of a quantum dot
can lead with some modest assumptions to a uniform distribution
of the signs of the tunneling matrix elements.

 Consider a quantum dot
which is defined in a 2D electron system 
by the circular harmonic confinement potential
\begin{eqnarray}
V_c(x,y) &=& \frac{1}{2}m_e \omega_0^2 (x^2 + y^2). 
\end{eqnarray}
Here $m_e$ is the electron's mass and $\hbar\omega_0$ is the energy
scale of the confining potential.
 The quantum dot is taken to be weakly coupled by tunneling to two leads
which can be described by one transport channel.
The tunneling between the two systems is assumed to be weak enough 
not to modify the electron's wave functions except for broadening 
the lineshapes. The Coulomb interaction in a dot is treated 
semiclassically.
With these assumptions 
the single-particle energy spectrum in the presence of perpendicular 
magnetic fields is given by the Fock-Darwin spectrum,
\begin{eqnarray}
E(n,m) &=& (2n + |m| +1) \hbar\Omega - \frac{1}{2}\hbar\omega_c, 
\end{eqnarray}
and the wave functions are 
\begin{eqnarray}
\label{waveftn}
\psi_{nm} (x,y) 
 &=& N_{nm} e^{-im\varphi} \rho^{|m|} 
   e^{-\rho^2/2} L_n^{|m|} (\rho^2).
\end{eqnarray}
Here $N_{nm} = \sqrt{\frac{n!}{\pi a^2 (n+|m|)!}}$ is a normalization
constant, $n$ is the nonnegative integer, and $m$ is an integer. 
The other variables in this equation are
$\rho = r/a$ with $a = \sqrt{\hbar/m_e\Omega}$, 
$\Omega = \sqrt{\omega_0^2 + \omega_c^2/4}$, and $\omega_c = eB/m_ec$. 
In the absence of magnetic fields, the energy spectrum reduces to
\begin{eqnarray}
E_N &=& (N+1) \hbar\omega_c, ~~~ N = 2n + |m|.
\end{eqnarray} 
Each energy level $E_N$ is degenerate with $D_N = N+1$, where $N=0,1,2,\cdots$.

 We now discuss the structure of the tunneling matrix elements. 
The tunneling barrier is assumed to be described by a real potential,
even in a magnetic field.
To study simply oscillating
phase-coherent transport, the two leads should accommodate 
only one transport channel.
Leads with more than one channel will show more complex patterns in
the AB oscillations.
 For a circularly symmetric quantum dot, the radial part of the wave
function determines the overall magnitude of the tunneling matrix elements
but plays no role in determining the their phase. The phase of the tunneling
matrix elements is determined solely by the angular part of the wave function.
For an ideal point contact between the two leads and the quantum dot
with one lead 
at $\varphi = 0$ and the other lead at $\phi =\pi$,
the relative phase between the left and right tunneling matrix elements
for the energy level labeled by $(n,m)$ is given by $e^{-im\pi} = (-1)^m$. 
The ideal point contact condition can be relaxed. 
If $\phi=0$ is replaced by $\phi \in [-\delta_1,\delta_1]$
and $\phi = \pi$ is replaced by $\phi \in [\pi-\delta_2, \pi+\delta_2]$,
then
for the constant tunneling over this range of $\phi$, the relative phase
is $(-1)^m$ times the signs of $\sin m\delta_2 / \sin m\delta_1$. 
As long as $\delta_{1,2} \ll \pi/2$, the relative phase is $(-1)^m$ for 
reasonable values of $m$. 
%

 In the absence of a magnetic field, any energy level 
consists of degenerate states with only even $m$ or only odd $m$.
The angular part of wavefunctions can be made real by the appropriate 
linear combinations of degenerate wavefunctions.
The relative signs are still given by $(-1)^m$.
The energy levels of even $m$ and odd $m$ alternate with increasing energy.
In a semiclassical approach to the Coulomb interaction, an electron added
to the quantum dot will occupy the single-particle states
which are unoccupied and lowest in energy. 
Until one single particle energy level $E_N$ is exhausted by $2(N+1)$ electrons
(including the spin degeneracy), the relative phases or signs will  
be equal 
so that the transmission zeros are generated in between the transmission peaks.
When the next higher energy level $E_{N+1}$ starts to be filled, the relative phase 
is now opposite to that of the lower energy level $E_N$. 
When $E_g$ is varied from the last electron filling at $E_N$ to the first
electron filling at $E_{N+1}$, there will not be a transmission zero. 
Successive transmission zeros start to appear again 
in the transmission amplitude while the energy level $E_{N+1}$ is filled by
electrons. 
The absence of the transmission zeros is realized only 
when going from $E_N$ to $E_{N+1}$.

 In the presence of a magnetic field, time-reversal symmetry is broken
and the wavefunctions cannot be chosen to be real. 
The relative signs are still $(-1)^m$ for the state $(n,m)$
if the tunneling barrier potential is real and the electron's wavefunction
in the one-channel leads are not modified in the presence of magnetic fields. 
The transmission phase will behave in the same way as for $B=0$
when the magnetic field is weak enough for level-crossing not to occur.
In a strong magnetic field, energy level-crossing starts to occur
and the distribution of the relative signs is modified as well.
We believe that keeping track of the relative signs is possible
experimentally just like monitoring the Coulomb peaks in the conductance.
When the magnetic field is strong enough that only the lowest Landau level
is filled, the energy levels will be ordered with an increasing
azimuthal quantum number $m$. For this case, the relative signs
will oscillate from level to level. However, the spin-degeneracy 
can double the oscillating period in the relative signs.

\section{Conclusion \label{sect_sum}}
 In this paper we considered several possible distributions of 
the relative signs in the tunneling matrix elements and studied the 
effects of these signs on the transmission zeros and phase. 
When the tunneling matrix elements are much smaller than the interlevel 
spacing or the distance between the Coulomb peaks, 
we find the following general rules governing the relation between the relative signs
and the transmission zeros or phase. 
 When the relative signs are opposite for the two neighboring energy levels, 
a transmission zero is absent in the energy interval bounded by the two levels.  
 When the relative signs are the same for two neighboring energy levels,
one transmission zero is generated inside the energy interval bounded by two levels.
Silva, et. al. \cite{signphase}
 studied a quantum dot with two energy levels
and found these same results. 
In our work we find that the above two rules are generally true
even in the presence of other energy levels in a dot,
as far as the linewidths are much smaller than the energy level spacing.  
Based on the above two rules, we can predict the behavior of the transmission 
zeros and phase from the distribution of the relative signs. 
A uniform distribution of signs within some energy interval
leads to transmission zeros in between any pair of conductance peaks.
The oscillating distribution of signs completely removes the transmission
zeros.

 When the strength of the tunneling matrix elements becomes larger
comparable to the interlevel spacing, the above simple rules are modified. 
One example is when one energy level is coupled much more strongly to 
the two leads than other levels.
When the strongly coupled level is broad enough to cover several neighboring 
energy levels, this broad level can act as the continuum band to the neighboring 
narrow levels and Fano interference occurs. 
The Fano interference results in an asymmetrical lineshape
 as well as the generation of new transmission zeros which are otherwise absent.

 Since we did not take into account in full length 
 the Coulomb interaction between electrons in the quantum dot, 
some care is needed to apply our theoretical results 
to the real quantum dots. 
In the Hartree approximation, the Coulomb energy mainly shifts the chemical potential
or determines the poles of the Green's function of a dot. 
When the tunneling matrix elements are so small that the Coulomb blockade
peaks are well separated and narrow, our theory will be relevant to the experiments.

 We have emphasized that the transmission zeros can be located by 
inspecting the Fano interference shape in $T_{AB}$  as a function of the gate voltage
and the magnetic field. 
The AB phase shifts the transmission zeros off the real-energy axis
and modulates the shape of $T_{AB}$. 
Depending on the existence or the absence of the transmission zeros in 
$T_Q$, the variation of $T_{AB}$ with the gate voltage is quite different.  
There is one experiment\cite{jpnissp} in which the linear conductance or $T_{AB}$ 
of the AB interferometer with an embedded quantum dot was measured. 
Based on our results [see Fig.~\ref{step} (c)], we can conclude that 
the signs of the tunneling matrix elements are flipped 
in the gate voltage range $-0.05V < V_g < 0V$ of Fig.~1(a) in the work\cite{jpnissp}
of Kobayashi, {\it et. al.}
Furthermore, all the energy levels covered by the gate voltage 
$-0.15V < V_g < -0.05V$ seem to have the same relative signs in the tunneling
matrix elements. 
The measured variation\cite{jpnissp} of the conductance $G_{AB}$ in Fig. 4(a)
is well reproduced by our $T_{AB}$'s for a uniform distribution of the 
relative signs.

 Our study shows that the asymmetrical lineshape of $T_Q(E_g)$
can result from the interference between the transport channels through the 
energy levels in a dot. 
Some experimental groups\cite{fanodotexp1,fanodotexp2}
have  observed an asymmetrical lineshape of $G_{Q} (E_g)$
as a function of the gate voltage.
This feature may be understood in terms of the interference
between multiple energy levels in a dot and the distribution of the relative
signs in the tunneling matrix elements.

 In summary, we studied the relationship between the transmission phase and 
the signs of the tunneling matrix elements.
The behavior of the transmission zeros and phase 
 is very sensitively dependent on distribution of the relative signs 
between the left and right tunneling matrix elements, and on their strength.
We suggest that the location of the transmission zeros can be 
identified by inspecting the shape of the conductance as a function of 
the gate voltage in the closed AB interferometers.

\acknowledgments
This work was supported 
in part by the National Science Foundation under Grant No. DMR 9357474, 
in part by the BK21 project,
and in part by grant No. 1999-2-114-005-5 from the KOSEF.

\appendix

\section{Complex Fano factor in the Aharonov-Bohm ring with an embedded quantum dot}
\label{sect_app}
 Asymmetric Fano lineshapes in the conductance have been observed in 
transport through the mesoscopic systems. 
\cite{jpnissp,fanodotexp1,fanodotexp2} 
The AB interferometer with an embedded quantum dot\cite{jpnissp} 
is one such system.  
In the main text we have already studied Fano interference patterns 
in the conductance of an AB ring.
In this appendix we are going to show that the complex Fano factor
is closely related to the complex transmission zeros in $t_{AB}$ 
or the transmission amplitude of an AB ring. 
The complex Fano factor is an oscillatory function of 
the magnetic fields threading through an AB ring.

 As Fano proposed\cite{Fano}, 
asymmetric Fano lineshapes arise from the interference between
the continuum energy spectrum and one discrete energy level. 
\begin{eqnarray}
\label{fanoeqn}
F(x) &=& F_{\infty} \frac{(x + q)^2}{x^2 + 1}. 
\end{eqnarray}
Here $q$ is the asymmetric Fano factor and is real in most cases. 
The Fano resonance is featured with a pair of one dip and one peak structures,
which is in direct contrast with the Lorentzian shape of a one peak structure.

To simplify the algebra and to make the physics clear, we consider 
a model quantum dot which is described by one resonant level. 
In this case Eq.~(\ref{tampring}) is simplified as 
\begin{eqnarray}
t_{AB} &=& \frac{ (\epsilon - \epsilon_d) \sqrt{T_0}
      + \overline{\Gamma} \sqrt{g} \cos \phi 
      + i \overline{\Gamma} \sqrt{g} \sin \phi  }  
    { \epsilon - \epsilon_d 
      + \overline{\Gamma} \sqrt{g\gamma} \cos \phi + i \overline{\Gamma} }.
\end{eqnarray} 
The overall phase is removed from $t_{AB}$ and $\epsilon_d$ is the resonant
energy level.
The direct tunneling probability $T_0$ is given by the expression
$T_0 = 4\gamma/(1+\gamma)^2$, where $\gamma = \pi^2 N_L N_R |T_{LR}|^2$,
$\overline{\Gamma} = (\Gamma_L + \Gamma_R)/(1+\gamma)$,
and $g = 4\Gamma_L\Gamma_R/(\Gamma_L + \Gamma_R)^2$ with
$\Gamma_{L/R} = \pi N_{L/R} |V_{L/R}|^2$. 
Writing $Z = \epsilon - \epsilon_d$, we can rewrite $t_{AB}$ 
in terms of the pole $Z_p$ and the zero $Z_z$ as
\begin{eqnarray}
t_{AB} &=& \sqrt{T_0} \frac{Z - Z_z} {Z - Z_p}. 
\end{eqnarray}
This zero-pole pair is given by the expressions, 
\begin{subequations}
\begin{eqnarray}
Z_p &=& -\overline{\Gamma} \sqrt{g\gamma} \cos \phi - i \overline{\Gamma}, \\
\label{dotzero}
Z_z &=& -\overline{\Gamma} \sqrt{\frac{g}{T_0}} [ \cos\phi + i \sin\phi].
\end{eqnarray}
\end{subequations}
Writing $x = (\epsilon-\epsilon_d)/\overline{\Gamma}$, the transmission
probability $T_{AB} = |t_{AB}|^2$ becomes
\begin{eqnarray}
T_{AB} 
 &=& T_0 \times \frac{|x + q(\phi)|^2}{(x + \sqrt{g\gamma}\cos\phi)^2 + 1},
\end{eqnarray}
where $q(\phi) = Z_z(\phi)/\overline{\Gamma}
 = \sqrt{\frac{g}{T_0}} e^{i\phi}$ 
is the complex asymmetric Fano factor. $T_{AB}$ is of the same form
as Eq.~(\ref{fanoeqn}) except that the asymmetric Fano factor is now 
a complex number depending on the AB phase $\phi$. 
The asymmetric Fano factor is proportional to
the transmission zero of the AB ring. 
The absolute magnitude of $q$, $|q| = \sqrt{\frac{g}{T_0}}$, is independent 
of the magnetic field in the case of one resonant level model.

 When the quantum dot is modeled by multiple resonant levels, 
the basic features of one resonant level model are not modified.
The complex transmission zeros in $t_{AB}$ become purely real when $\phi = n\pi$
with an integer $n$. Though the imaginary part of the zeros in $t_{AB}$
is still sinusoidal as a function of $\phi$, its functional form is 
not given by the form $\sin\phi$.\cite{tskim}
The absolute value of the complex Fano factor in this case
is weakly dependent on magnetic field.

\end{document}